\documentclass{PoS}
\usepackage{amssymb,epsfig,amscd}
\usepackage{amsmath}
\usepackage{amsfonts}
\usepackage{amssymb}
\usepackage{graphicx}%
\title{(Oscillating) non-exponential decays of unstable states}

\ShortTitle{(Oscillating) non-exponential decays of unstable states}

\author{\speaker{Francesco Giacosa}\\Institut f\"ur Theoretische Physik, Goethe Universit\"at, Max-von-Laue-Str. 1,
60438 Frankfurt am Main, Germany \\E-mail: giacosa@th.physik.uni-frankfurt.de}
\author{Giuseppe Pagliara\\Dipartimento di Fisica, Universit\'a di Ferrara and INFN Ferrara, via Saragat
1, 44100 Ferrara, Italy\\E-mail: pagliara@fe.infn.it }

\abstract{We discuss deviations from the exponential decay law which occur
when going beyond the Breit-Wigner distribution for an unstable state. In particular, we
concentrate on an oscillating behavior, remisiscent of the Rabi-oscillations, in the short-time
region. We propose that these oscillations can explain the socalled GSI anomaly, which
measured superimposed oscillations on top of the exponential law for hydrogen-like nuclides decaying
via electron-capture. Moreover, we discuss the possibility that the deviations from the Breit-Wigner
in the case of the GSI anomaly are (predominantely) caused by the interaction of the unstable 
state with the measurement apparatus. The consequences of this scenario, such as the non-observation
of oscillations in an analogous experiment perfromed at Berkley, are investigated.}

\FullConference{50th International Winter Meeting on Nuclear Physics\\
23-27 January 2012 \\
Bormio, Italy}

\begin{document}

\section{Introduction}

The decay law of unstable systems plays a crucial role in Physics: the
electromagnetic decays of atoms, the decays of radioactive nuclei, of hadronic
resonances and of the Standard Model particles such as the weak interaction
bosons and the Higgs boson, are all described by the well known exponential
decay law. Once the decay rate $\Gamma$ is calculated from the microscopic
interactions or measured in experiments, the decay law is simply given by
$p(t)=e^{-\Gamma t}$, where $t$ is the time after the preparation of the
unstable state and $p(t)$ represents the survival probability.

On the other hand, it is a fact that both in Quantum Mechanics \cite{ghirardi}
and in Quantum Field Theory \cite{Giacosa:2010br,Pagliara:2011hh} a pure
exponential decay law is not obtained: deviations from the exponential law are
present at times very close to the initial preparation time $t=0$ and at very
late times, while at \textquotedblleft intermediate\textquotedblright\ times
the exponential law represents a very good approximation. In particular, at
late times the decay law follows a power-law, which is however very difficult
to observe experimentally because it occurs at times for which the survival
probability is already vanishingly small. On the other hand, the deviations at
small times occur within a very short time scale, for instance $10^{-15}$ s
for the electromagnetic decays of an excited hydrogen atom
\cite{pascaziohydrogeno} and even shorter for hadronic decays
\cite{Giacosa:2010br}. It is thus experimentally very challenging to observe
such deviations and to confirm the predictions of the theory. Only in 1997
cold atom experiments allowed to clearly observe for the first time deviations
from the exponential decay law of unstable systems (via tunneling of atoms out
of a trap) \cite{raizen}. In particular, this experiment has shown that the
survival probability at small times is not exponential, but it is rather a
Gaussian, i.e. the derivative of $p(t)$ goes to zero at times close to the
initial time, $p^{\prime}(0)=0$. In turn, this behavior allows for a quite
peculiar modification of the decay law induced by measurements: when pulsed
measurements on the system (inducing a collapse of the state into the original
undecayed state) are performed during the non-exponential regime, one can
obtain a slower or faster decay of the system depending on the frequency of
the measurements. Those two effects, called Quantum Zeno and Anti-Zeno effects
theoretically predicted in Refs.
\cite{1977JMP....18..756M,2002quant.ph..2127F,1996IJMPB..10..247N,2008JPhA...41W3001F}%
, have been then observed in the same experiment which has proven the
existence of non-exponential decays \cite{raizen1}. This experimental success
triggered a new interest of the physics community on the topic of deviations
from the exponential decay law, not only because it represents a new and deep
confirmation of the predictions of quantum theory, but also because it opens
the possibility to engineer the decay of unstable quantum systems, see for
instance Ref. \cite{kofman}. Also, the general theory of measurement in
quantum mechanics, which is still a quite active area of research, benefits
from these experimental results \cite{koshino}.

In 2008, an experiment at the Storage Ring of the GSI facility of Darmstadt
has reported the observation of non-exponential decays of hydrogen-like ions
which decay via electron capture \cite{Litvinov:2008rk}. Quite remarkably, the
survival probability shows an exponential decay with superimposed
oscillations. These data stimulated many discussions and many different
possible explanations have been proposed. Presently, there is no accepted
theoretical explanation of this phenomenon and, more important, an
experimental confirmation of the results is still lacking. By assuming that
the phenomenon observed at GSI is real, we present a possible explanation in
terms of \textquotedblleft standard\textquotedblright\ quantum mechanical
effects \cite{noigsi}. Moreover, we present further consequences of our
explanation which can be proved or disproved in the near future. We also
discuss the results obtained at the Berkeley Lab where no oscillations for the
decays of the same nuclei have been observed \cite{vetter}.

\begin{figure}[ptb]
\bigskip\begin{centering}
\epsfig{file=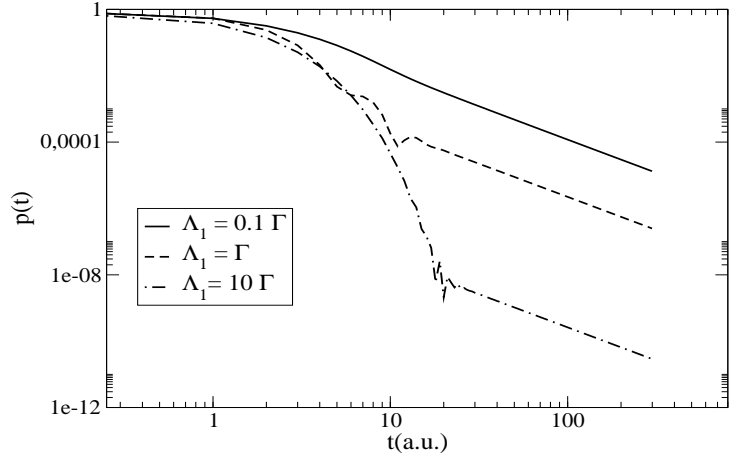,height=9.5cm,width=6cm,angle=-90}
\caption{Survival probability for different values of the low energy cutoff $\Lambda_1$.
$\Gamma=1$, $\Lambda_2=200 \Gamma$ (a.u.). The exponential decay law turns into a power law
at large times. }
\end{centering}
\end{figure}

\section{A phenomenological approach to unstable states}

The standard empirical approach to the decay of unstable states is to assume
that the decay rate $\Gamma$, i.e. the number of decays per unit time, is
constant and does therefore not depend from the \textquotedblleft
history\textquotedblright\ of the unstable system, in particular it does not
depend on the time of \textquotedblleft preparation\textquotedblright\ (in a
quantum mechanical meaning) of the unstable state. This immediately leads to
the exponential survival probability $p(t)=e^{-\Gamma t}$.

Denoting the energy distribution function (alias the spectral function) of the
unstable state with $d(x)$ \cite{ghirardi}, the survival amplitude $a(t)$ is
the Fourier transform of $d(x)$, $a(t)=\int_{-\infty}^{\infty}\mathrm{dx}%
d(x)e^{-ixt},$ and the survival probability is just given by $p(t)=\left\vert
a(t)\right\vert ^{2}$. The \textquotedblleft empirical\textquotedblright%
\ exponential decay law is theoretically justified, if we assume that $d(x)$
is a Breit-Wigner distribution $d(x)\rightarrow d_{BW}(x)=\frac{1}{2\pi}%
\frac{\Gamma}{(x-M)^{2}+\Gamma^{2}/4}$ , where $M$ is the mass of the unstable
state (i.e. its energy in the rest frame) and $\Gamma$ its decay width. Note
that $a(0)=1$ (the state is prepared at the instant $t=0$ with unit
probability, $p(0)=1$). When calculating the Fourier transform, the integral
gets only the contribution of the simple pole located at $x_{pole}%
=M-i\Gamma/2$ leading to $a_{BW}(t)=e^{-iMt}e^{-\Gamma t/2}$ and thus
$p_{BW}(t)=e^{-\Gamma t}$.

The exponential law works astonishingly well when compared to the experiments.
However, there are two evident problems in assuming a Breit-Wigner spectrum:
(i) It does not allow for the existence of a minimum of energy (threshold for
the decay), i.e. it corresponds to an Hamiltonian unbound from below. (ii) The
behavior of the Breit-Wigner at large energies is such that, while the
normalization can be imposed (and thus unitarity), all the momenta of the
distribution, including also the average energy of the unstable state,
diverge. We need to cure these two problems in order to build a physically
motivated distribution function. In Quantum Field Theory, once the interaction
Hamiltonian between the unstable state/particle and the decay products is
known, the spectral function is proportional to the imaginary part of the
dressed propagator, i.e. the one obtained by the resumming all the loops, see
Ref. \cite{Giacosa:2007bn}. This procedure allows to correctly describe the
spectral function in the whole energy spectrum. In particular, there appears a
threshold which then, in the survival probability, regulates the decay law of
the system at large times, for which a power law is realized. The high energy
behavior of the spectral function, which controls the survival probability at
small times, represents unfortunately a much more complicated problem. All
field theories are valid until a cutoff of energy where some new physical
ingredients enter, for instance the Planck energy scale in particle physics.
Following this reasoning, we construct here the simplest phenomenological
model for the spectral function and, as a consequence, for the survival
probability: we assume a Breit-Wigner spectrum corrected by two cuts in the
energy, a low energy cut $\Lambda_{1}$ and a high energy cut $\Lambda_{2}$:%
\begin{equation}
a(t)=N\int_{M-\Lambda_{1}}^{M+\Lambda_{2}}\mathrm{dx}\frac{\Gamma}%
{(x-M)^{2}+\Gamma^{2}/4}e^{-ixt}\text{ ,}\label{oura}%
\end{equation}
where $N$ is such that $a(0)=1.$ For such distribution, obviously all the
momenta are finite (in particular the average energy of the unstable state).
We show in Figs. 1 and 2 the effect of varying these two parameters on the
survival probability $p(t)$ and on the decay rate as a function of time. The
quantity $h(t)$ defined as $h(t)=-dp(t)/dt$ ($h(t)dt$ represents the
probability that the unstable state decays in the time interval between $t$
and $t+dt$).

As expected, the exponential behavior dominates over a very long time scale
but deviations are clearly present. The low-energy cutoff $\Lambda_{1}$
regulates the time after which the exponential law turns into a power law (in
our approach the index of the power law cannot be adjusted, one should
introduce another parameter which fixes how fast the decay rate falls to zero
at threshold). When varying the high-energy cutoff $\Lambda_{2}$, the survival
probability remains basically very similar to an exponential law (see Ref.
\cite{noigsi} and figures therein), but very interesting features emerge in
the behavior of $h(t)=-p^{\prime}(t)$. The high energy cutoff regulates the
behavior of the decay probability at small times after the preparation of the
system and manifests itself (in the Fourier transform) as an oscillation
superimposed to the exponential decay law. The larger the cutoff, the larger
is the frequency of the oscillation and the smaller is the amplitude of the
oscillation as one can notice in Fig. 2 (where, for simplicity, the choice
$\Lambda_{1}=\Lambda_{2}$ has been made). The physical interpretation of this
phenomenon is quite natural: $\Lambda_{2}$ determines the bandwidth of the
continuum of states into which the unstable state can decay. The plots shown
in Fig. 2 interpolate between the pure exponential decay which occurs in
presence of a large bandwidth continuum and a pure oscillating probability
which occurs in a system of two discrete levels ($\Lambda_{2}$ very close to
the average energy of the unstable system) where Rabi oscillations are obtained.

The physical origin of the cutoff(s) can be twofold: there can be
\textquotedblleft natural\textquotedblright\ cutoffs determined by the
microphysics of the interaction of the unstable state and its decay products
(as the low-energy threshold and the high-energy cutoff mentioned above), but
there could be an \textquotedblleft experimental\textquotedblright\ cutoff,
which is caused by the interaction of the unstable system with the
experimental apparatus that measures the decay. As we will discuss in the
following, in the case of the GSI anomaly the experimental cutoff must
dominate (i.e. it is the smaller one).

\bigskip

\begin{figure}[ptb]
\bigskip\begin{centering}
\epsfig{file=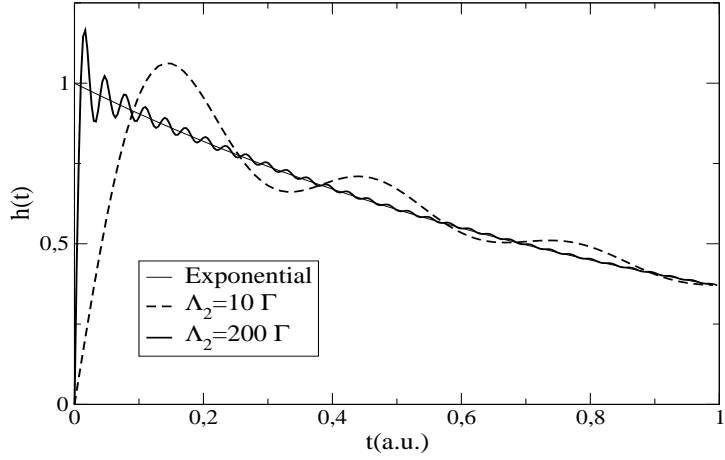,height=9.5cm,width=6cm,angle=-90}
\caption{Decay rate as a function of time $ h(t)$ for different choices
of the high energy cutoff $\Lambda_2$. $\Gamma=1$, $\Lambda_1=\Lambda_2$ (a.u.). Also the
exponential decay rate is shown for comparison.}
\end{centering}
\end{figure}

\section{The peculiar case of the GSI anomaly}

The interesting properties of the decay law explained before, especially the
oscillating behavior of $h(t)$ shown in Fig. 2 emerging for \textquotedblleft
small\textquotedblright\ values of the cutoff(s) (one or two orders of
magnitude larger than $\Gamma$), \emph{might} have been already observed in
experiments. In particular, we want to point to what has been named the
\textquotedblleft GSI anomaly\textquotedblright\ seen at the heavy ions
storage ring at the GSI facility of Darmstadt. In 2008 Ref.
\cite{Litvinov:2008rk} reported the observation of non-exponential decays of
Hydrogen-like ions $^{140}$Pr and $^{142}$Pm in the electron-capture reactions
of the form:
\begin{equation}
M\rightarrow D+\nu_{e} \label{ec}%
\end{equation}
where $M$ denotes the `mother state' (i.e. the unstable $H$-like nuclide
$^{140}$Pr or $^{142}$Pm) and $D$ denotes the `daughter state' (i.e. the
nuclei $_{58}^{140}$Ce and $_{60}^{142}$Nd, respectively).

Calling $N(t)$ the number of unstable particles at the instant $t$, it has
been found that $dN/dt$ does $\emph{not}$ follow a simple exponential law. The
experimental points were fitted with superimposed oscillations:
\begin{equation}
\frac{dN_{dec}}{dt}=-\frac{dN}{dt}\propto e^{-\lambda t}(1+a\cos(\omega
t+\phi))\text{ ,} \label{fitgsi}%
\end{equation}
where $dN_{dec}/dt$ represents the number of decay per time (see Fig. 3-5 of
Ref. \cite{Litvinov:2008rk}). These results stimulated the theoretical
modelling of this phenomenon, but the origin of these oscillations is not yet
clear: explanations of the observed experimental data by invoking neutrino
oscillations, neutrino spin precession and quantum beats seem indeed not to be
satisfactory, see Refs. \cite{Cohen:2008qb,gal,Merle:2010qq,Wu:2010ke} and
refs. therein.

In Ref. \cite{noigsi} we have put forward an interpretation of the GSI results
based only on Quantum Mechanics: Following the discussions of Sec. II, we
assumed that the mass distribution of the mother state is not a pure
Breit-Wigner. In doing the calculations the survival probability amplitude of
Eq. (\ref{oura}) with $\Lambda=\Lambda_{1}=\Lambda_{2}$ has been used. We have
shown that for $^{142}$Pm the cutoff $\Lambda\simeq32\Gamma\simeq
0.5\cdot10^{-15}$ eV, where $\Gamma=0.0224$ s$^{-1}$ is the decay width of the
state, gives rise to oscillations which are qualitatively similar to those
measured in Ref. \cite{Litvinov:2008rk}.

The required value of the cutoff $\Lambda$ is very small. An intrinsic origin
of this cutoff based on QED and QCD fundamental interactions, on the line of
Ref. \cite{pascaziohydrogeno}, seems very improbable in this case. A more
promising direction consists in assuming that the cutoff originates from the
interaction of the unstable system with the measuring apparatus. Indeed, the
experiment performed at the GSI storage ring is unique. After their
\textquotedblleft creation\textquotedblright, the unstable Hydrogen-like
$_{59}^{140}$Pr and $_{61}^{142}$Pm ions are stored in a ring equipped with a
Shottky Mass spectrometer which measures the frequency of rotation of the ions
inside the ring. This frequency depends on the charge to mass ratio of the
ions. When the reaction (\ref{ec}) takes place, the charge to mass ratio, and
thus the rotation frequency, change. In this way the experimentalists at GSI
can monitor the decay of these unstable systems few seconds after their
preparation and for a period of a couple of minutes. Some important features
of this experiment are worth to be mentioned: While the ions need $\sim0.5$
$\mu$s to complete a turn in the storage ring, their frequency of rotation
within the frequency spectrum is identified within averagely $\Delta
t_{resolution}\simeq200$ ms. This means that the `measurement' of the state of
the unstable ions does not occur at every turn\footnote{Another important
experimental limitation concerns the time interval which lasts between the
disappearance of the frequency of the mother ion and the appearance of the
frequency of the daughter ion in the frequency spectrum. This time interval of
$900$ ms and $1200$ ms (for $^{140}$Pr and $^{142}$ Pm respectively) is
related to the cooling of the stream before it can be identified by the mass
spectrometer.}. This measurement is clearly not an `ideal measurement' in the
quantum mechanical sense, according to which the collapse of the wave function
occurs instantaneously as soon as the wave function of the unstable system
interacts/overlaps with the measurement apparatus (the projection postulate of
Quantum Mechanics). We connect the cutoff $\Lambda$ entering into the
expression (\ref{oura}) to the precision of the experiment through the
time-energy uncertainty relation:
\begin{equation}
\Lambda\sim\frac{1}{\Delta t_{resolution}}\sim10^{-15}\text{ eV .}\label{ldt}%
\end{equation}
This number is remarkably close to the value (mentioned above) needed to
describe the oscillation seen in the GSI experiment. It seems therefore that
the possibility of an apparatus-induced cutoff is viable and deserves further
discussion. Indeed, the fact that the measurement itself can modify the decay
law of an unstable system has been already experimentally proven by the
observation of the quantum Zeno and Anti-Zeno effects \cite{raizen1}. To our
knowledge, the first theoretical work on this problem is Ref.
\cite{degasperis} \textquotedblleft Does the Lifetime of an Unstable System
Depend on the Measuring Apparatus\textquotedblright\ and recently a new
interest on this issue has grown, see Refs. \cite{kofman,koshino} and refs.
therein. In Ref. \cite{degasperis} it is analyzed how the decay of an unstable
state into two particles is modified by the measuring apparatus such as a
bubble chamber. A length scale $R$ is introduced and named \textquotedblleft
localization radius\textquotedblright,\ which corresponds to the distance
between the decay products beyond which the experimental apparatus can
ascertain whether the system has decayed or not. In this scheme, the following
formula for $p(t)$, formally identical to Eq. (\ref{oura}) for $\Lambda
=\Lambda_{1}=\Lambda_{2}=w/2$, has been obtained by studying the dependence of
the lifetime of the unstable state from the experimental apparatus:
\begin{equation}
p(t)=N\left\vert \int_{E_{R}-w/2}^{E_{R}+w/2}dE\frac{e^{-iEt}}{(E-E_{R}%
)^{2}+\gamma^{2}/4}\right\vert ^{2}\text{ ,}\label{dg}%
\end{equation}
where $N$ is a normalization constant, $E_{R}$ is the average energy (i.e. the
mass $M$) of the unstable state, $\gamma$ the width of the state (i.e.
$\Gamma$) if it is not disturbed by the measurement, and the cutoff $w$ is
proportional to the ratio of relative velocity $v$ of the two decay products
and the localization radius $R$, $w=v/R$. The interpretation of $w$ in the
case of Ref. \cite{degasperis} is quite clear: it is related to the time
needed by the measuring apparatus to destroy the correlation between the
unstable state and its decay products and it is thus of the same type of Eq.
(\ref{ldt}). As discussed in that paper, for the typical measuring apparatuses
in particle physics (as for instance the bubble chamber) $w$ is very large:
$10^{17}$-$10^{22}$ s${^{-1}}$ and therefore the exponential decay is obtained
to a very good approximation in most of the cases. The arguments in Ref.
\cite{degasperis} leading $w$ are analogous to the emergence of a cutoff
$\Lambda=w/2$ in our case.

Indeed, one should go beyond these qualitative considerations and build a
detailed theoretical model for the interaction between the unstable ions and
the measuring apparatus also in the case of the GSI experiment, but this
represents a quite demanding effort which is left for future work. Moreover,
it will be also important to study in detail the effect of the `collapse' of
the wave function in the case of the GSI experiment. As discussed in Ref.
\cite{hotta,delgado}, the non-occurrence of the Zeno effect (and therefore the
clock is not reset at each measurement) would assure that the quantity
measured in the experiment coincides (up to a normalization) with the function
$h(t)=-p^{\prime}(t)$. These issues go at the very heart of Quantum Mechanics:
in this sense, the GSI experiment could represent a wonderful way to directly
investigate them.

As a next step we list the predictions and consequences which hold in the
framework of our proposed interpretation.

(i) The curve $h(t)=-p^{\prime}(t)$: Our theoretical function $h(t)$ (which
represents the decay probability per unit time and unit ion and it is thus
proportional to $dN_{dec}/dt$) evaluated starting from Eq. (\ref{oura}) shows
some peculiar differences w.r.t. the experimental fitting curve of Eq.
(\ref{fitgsi}). Our $h(t)$ vanishes for short times (a general feature due to
the fact that $p^{\prime}(0)=0$), the first peak is more pronounced than the
others and the oscillations are damped faster than the fitting curve in Eq.
(\ref{fitgsi}), see Fig. 2 and the detailed discussion and figures in Ref.
\cite{noigsi}.

(ii) $\beta^{+}$ decay channel: The $H$-like ions under study at the GSI do
not decay only via the electron-capture mechanism of Eq. (\ref{ec}), but decay
(in both cases sizably) via a $\beta^{+}$ decay: $M\rightarrow D^{\prime
}+e^{+}+\nu_{e},$ where $D^{\prime}$ refers to the $H$-like daughter state for
this process. In the case of the $\beta^{+}$ decay a positron is emitted which
is absorbed by the environment extremely fast. Thus, for the $\beta^{+}%
$-channel the corresponding cutoff turns out to be much larger than $10^{-15}$
eV: the deviations from the exponential decay law are very small and thus
unobservable in this channel (see Fig. 2 to `see' the effect of an increased
cutoff). This discussion is also useful to clarify the following point: at
variance with the positron, in the electron-capture decay of Eq. (\ref{ec})
the emitted neutrino does not interact with matter and is therefore not
responsible for the determination of a time scale. For a mathematical
description of the two-channel case we refer to Ref. \cite{2canali}. A
detailed study of the two-channel problem using the formalism of Ref.
\cite{2canali} is also part of our outlook.

(iii) Berkeley-experiment: In the experiment performed at the Berkeley Lab
\cite{vetter} \emph{no} oscillations in the decay law for $^{142}$Pm in
relation to the `same' process of Eq. (\ref{ec}) have been observed. As
already noticed in Ref. \cite{faestermann}, there are peculiar differences
from the GSI experiment w.r.t. the Berkeley one: in the latter, the atoms are
not ionised and are inside a lattice, thus also phonons are emitted in the
final state. However, in the framework of our interpretation, the crucial fact
is that very soon after the electron capture of Eq. (\ref{ec}), a $K$-shell
vacancy is formed and a photon is very soon emitted. Thus, just as in the
previous case, $\Delta t$ is much shorter and, conversely, the cutoff is much
larger in the Berkeley-experiment: $\Lambda_{Berkeley}\gg\Lambda.$ The
oscillations have a too small amplitude and period and cannot be observed.
Moreover, the absence of oscillations at the Berkeley experiment is a further
strong argument against an intrinsic cutoff emerging out of microscopic form factor.

(iv) Independence of the period and amplitude on the employed $H$-like ion: In
the framework of our interpretation, the cutoff $\Lambda$ is almost uniquely
related to the measurement process and is therefore independent of the
employed mother nuclide. Thus, the period and the amplitude of the
superimposed oscillations, which are controlled by the cutoff only, are also
expected to be comparable: this is indeed the case of the two ions studied in
Ref. \cite{Litvinov:2008rk}. Notice that the same cutoff of Eq. (\ref{ldt})
for both ions $^{140}$Pr and $^{142}$Pm corresponds to quite different ratios
of $\Lambda/\Gamma,$ which are $\sim32$ and $\sim470$ respectively. It is
interesting that the measurement-induced cutoff can explain naturally these
quite different ratios.

(v) Repetition of the experiment. If the GSI experiment is performed with an
improved time resolution, we expect that the corresponding cutoff increases,
see Eq. (\ref{ldt}), and thus the period and the amplitude of the oscillations
decrease, see Fig. 2 for a numerical example.

Finally, it should be stressed that, while the here described qualitative
features are general, a quantitative analysis should go beyond the simple
formula of Eq. (\ref{oura}). This will be possible once that, as mentioned
above, a detailed study of the interaction of the system as a whole (unstable
state plus measurement) will be undertaken.

\section{Conclusions}

In this work we have described deviations from the exponential decay law when
the energy distribution is not a Breit-Wigner function. In particular, we have
studied a modified energy distribution in which cutoffs on the left and on the
right sides of the peak have been introduced. We have proposed that the
oscillations seen in Ref. \cite{Litvinov:2008rk} in the electron-capture decay
of $H$-like ions may originate from a similar modification of the energy
distribution of the mother state \cite{noigsi}. Inspired by Ref.
\cite{degasperis}, we have linked through the time-energy uncertainty relation
the physical origin of the cutoff $\Lambda$ to the time uncertainty of the
measuring apparatus at GSI. It is quite remarkable that the cutoff obtained in
this way, see Eq. (\ref{ldt}), is of the same order needed to obtain the time
modulation of $7$ s measured in Ref. \cite{Litvinov:2008rk}.

We have analyzed the consequences of our proposal: very much suppressed
oscillations in the $\beta^{+}$ decay-channel because of a much larger cutoff,
which makes them unobservable: the standard decay law holds here; similarly,
suppressed oscillations (and thus exponential decay law) in the
electron-capture decay channel at the Berkeley experiment; a mild dependence
of the period and amplitude on the unstable ion; more suppressed oscillations
when the GSI experiment is repeated with an increased time resolution (period
and amplitude decrease).

As an outlook for future works we mention the precise modelling of the
measurement procedure and the detailed study of the two-channel problem.

\bigskip

\acknowledgments G.P. acknowledges financial support from the Italian Ministry
of Research through the program Rita Levi Montalcini. F. G. acknowledges
useful discussions with F. Bosch, Y. Litvinov, N. Winckler, S. Bishop, and D. Shubina.

\end{document}